\documentclass[twocolumn]{emulateapj}
\usepackage{amsmath}
\usepackage{rotating}
\usepackage{hyperref}
\usepackage{graphicx}
\usepackage{epsfig}
\usepackage{color}
\def\galaxiespprox{\mathrel{\vcenter{\offinterlineskip \hbox{$>$}
    \kern 0.3ex \hbox{$\sim$}}}}
\def\lapprox{\mathrel{\vcenter{\offinterlineskip \hbox{$<$}
    \kern 0.3ex \hbox{$\sim$}}}}
\newcommand{\beq}{\begin{equation}} 
\newcommand{\eeq}{\end{equation}}

\def\eps{\epsilon}
\def\epsw{\eps_{\rm w}}

\def\Sig15{\Sigma_{1.5}}

\def\kms{\rm\thinspace km~s^{-1}}

\def\kpc{{\rm\thinspace kpc}}
\def\pch{{\rm\thinspace pc} \thinspace h^{\rm -1}}
\def\Msun{\hbox{$\thinspace \rm M_{\odot}$}}
\def\Msunh{\hbox{$\thinspace  \it h^{\rm -1} \rm \thinspace M_{\odot} $}}

\def\re{r_{\rm e}}
\def\reff{r_{\rm eff}}

\def\Mstar{M_{\ast}}

\def\Mdotacc{\dot{M}_{\rm acc}}

\def\Mdotoutf{\dot{M}_{\rm outf}}

\def\strom{Str$\rm \ddot{o}$mgren$\thinspace$}

\begin{document}
\shortauthors{Choi et al.}
\shorttitle{AGN feedback and quiescent galaxies}
\title{The Role of Black Hole Feedback on Size and Structural Evolution in Massive Galaxies}
\author{Ena Choi\altaffilmark{1,2}, 
              Rachel S. Somerville\altaffilmark{2,3},
              Jeremiah P. Ostriker\altaffilmark{1,4},
              Thorsten Naab\altaffilmark{5},
              Michaela~Hirschmann\altaffilmark{6}
}
\affil{ $^1$Department of Astronomy, Columbia University, New York, NY 10027, USA \\    
        $^2$Department of Physics and Astronomy, Rutgers, 
        The State University of New Jersey,\\
        136 Frelinghuysen Road, Piscataway, NJ 08854, USA\\
        $^3$Simons Center for Computational Astrophysics, New York, 
        NY 10010, USA \\
        $^4$Department of Astrophysical Sciences, Princeton University, Princeton, NJ 08544, USA\\
        $^5$Max-Planck-Institut f\"ur Astrophysik,
        Karl-Schwarzschild-Strasse 1, 85741 Garching, Germany \\
        $^6$Sorbonne Universite's, UPMC-CNRS, UMR7095, 
         Institut d' Astrophysique de Paris, F-75014 Paris, France
  \\ 
  \texttt{ena.choi@columbia.edu} \\}

\begin{abstract}
We use cosmological hydrodynamical simulations to investigate the role 
of feedback from accreting black holes on the evolution of sizes, 
compactness, stellar core density and specific star-formation of massive 
galaxies with stellar masses of $ \Mstar  >  10^{10.9} \Msun$. 
We perform two sets of cosmological zoom-in simulations of 30 halos to $z=0$:
(1) without black holes and Active Galactic Nucleus (AGN) feedback and 
(2) with AGN feedback arising from winds and X-ray radiation. 
We find that  AGN feedback can alter the stellar
density distribution, reduce the core density within the central 1 kpc by 
0.3 dex from $z=1$, and enhance the size growth of massive galaxies. We 
also find that  galaxies simulated with AGN feedback evolve along 
similar tracks to those characterized by observations in specific star 
formation versus compactness. 
We confirm that AGN feedback plays an important role in transforming 
galaxies from blue compact galaxies into red extended galaxies 
in two ways: (1) it effectively quenches the star formation, transforming 
blue compact galaxies into compact quiescent galaxies and (2) it also 
removes and prevents new accretion of cold gas, shutting down in-situ 
star formation and causing subsequent mergers to be gas-poor or mixed. Gas 
poor minor mergers then build up an extended stellar envelope. AGN 
feedback also puffs up the central region through the fast AGN 
driven winds as well as the slow expulsion of gas while the black hole is 
quiescent. Without AGN feedback, large amounts of gas accumulate in the 
central region, triggering star formation and leading to overly massive blue 
galaxies with dense stellar cores.
\end{abstract}

\section{Introduction}\label{intro}
In the local universe, galaxies show a bimodal color distribution: red
quiescent galaxies (QGs) with old stellar populations, and young, blue, and
star-forming galaxies (SFGs) \citep[e.g.][]{2003MNRAS.341...54K,
2004ApJ...608..752B}. The color bi-modality has been observed
at early epochs up to $z \sim 2 \-- 3$ \citep{2007ApJ...665..265F,
2009ApJ...706L.173B}. Quiescence seems to be strongly 
correlated with structural properties such as galaxy size and concentration, 
at all epochs since $z \sim 3$ --- at a {\it given} stellar mass,
QGs typically have smaller half-light radii, and denser cores, than SFGs 
\citep{2003MNRAS.343..978S,2010ApJ...713..738W,2011ApJ...738..106W,
2012ApJ...753..167B,2013ApJ...776...63F,2014MNRAS.440..843O,
2014ApJ...788...28V,2017MNRAS.465..619B,2017MNRAS.472.2054P}. 
The fraction of quiescent galaxies in the SDSS is found to be correlated 
with the mass and velocity dispersion of the bulge 
\citep{2014MNRAS.441..599B,2016MNRAS.462.2559B}.

The sizes of QGs are observed to be much smaller at fixed stellar 
mass at $z \sim 2$ compared to local elliptical galaxies 
\citep{2004ApJ...600L.107F,2004ApJ...604..521T,
2005ApJ...626..680D,2007MNRAS.374..614L,2007ApJ...671..285T,
2007MNRAS.382..109T,2008ApJ...687L..61B,2008A&amp;A...482...21C,
2008ApJ...688..770F,2008ApJ...677L...5V,2010ApJ...714L.244S,
2014ApJ...788...28V}. 
These compact QGs (cQGs), also known as ``red nuggets'' 
\citep{2009ApJ...695..101D}, have sizes of about $R_{eff} = 1 \kpc$, about 
four times smaller than local QGs of the same mass. Moreover, the 
number density of compact galaxies is much lower in the local universe 
than in the past \citep{2010ApJ...720..723T,2011ApJ...743...96C}, leading 
to much speculation as to the fate of compact galaxies, and how these 
different populations of compact, extended, SF, and Q galaxies are related 
to one another at different cosmic epochs.

\citet{2013ApJ...765..104B} suggested that massive galaxies evolve 
through a characteristic track in the plane of sSFR versus compactness 
(defined as $\Sig15 \equiv m_*/\re^{1.5}$, where $m_*$ is the galaxy 
stellar mass and $r_e$ is the half-light radius; see their Figure 6): 
beginning at around redshift $z\sim 2.5$--3, compact SFGs (cSFGs) 
begin to quench, building up a population of compact quiescent 
galaxies (cQG). Below $z\sim 1$, galaxies increase in mass and (even 
more so) in size, to form the extended quiescent galaxies that are 
common today. The results of \citet{2013ApJ...765..104B} furthermore 
suggest that the quenching time is relatively short (300 Myr to 1 Gyr) 
\citep[see also][]{2017MNRAS.472.2054P}. \citet{2014ApJ...791...52B} 
showed that the radii and stellar mass surface densities of compact 
SFGs (cSFGs) quantitatively matched those of cQGs, supporting the 
picture in which cSFGs rapidly quench into cQGs.  

A similar picture was suggested by \citet{2015ApJ...813...23V}, who 
also noted the concurrent decline of the cSFG population and rise of 
the cQG population. They proposed a toy model, in which star forming 
galaxies evolve along parallel tracks in the size-mass plane with 
$\Delta r\propto\Delta m^{0.3}$, until they reach a stellar density 
or velocity dispersion threshold and quench their star formation. After 
quenching, galaxies evolve along a steeper track in the size-mass 
plane, with $\Delta r\propto\Delta m^{2}$ as observed 
\citep{2010ApJ...709.1018V,2013ApJ...766...15P}. 

These observationally inferred formation scenarios pose several major 
questions to theoretical galaxy formation models: Which physical 
processes drive the characteristic evolution of the structural properties 
of star forming and quiescent galaxies?
What physical process is responsible for the quenching of star formation? 
Are these two sets of processes causally connected, or the result of a 
common cause, or do they occur together only coincidentally? What 
causes the change in slope of the evolutionary path in the size-mass 
plane, or put another way, why do quiescent galaxies grow so much 
more rapidly in radius relative to their mass growth than star forming 
galaxies?

Several potential reasons for the early development of compact 
galaxies have been addressed in many papers, such as intense 
starbursts in galactic nuclei triggered by gas-rich major mergers 
\citep{1991ApJ...370L..65B,2006ApJS..163...50H,
2015MNRAS.449..361W}, strong gas inflows within the disk driven 
by violent disk instabilities \citep{2001ApJ...553..174G,
2009ApJ...703..785D}, and early dissipative assembly 
\citep{2010ApJ...725.2312O}. Different authors have applied different
labels to this phase of galaxy formation, but all pictures involve
rapid inflow of gas into central star forming regions 
\citep[see reviews by][]{2015ARA&amp;A..53...51S,
2017ARA&amp;A..55...59N}. Some works have suggested a causal 
connection between these dissipative processes (``compaction") and 
quenching via a combination of gas exhaustion, and stellar and AGN 
driven winds \citep{2014MNRAS.438.1870D,2015MNRAS.450.2327Z}. 
Observationally, strong galactic outflows are often associated with 
high star formation and gas densities 
\citep[e.g.][]{2012ApJ...755L..26D}.

The $\Lambda$ Cold Dark Matter ($\Lambda$CDM) hierarchical 
structure formation model ubiquitously predicts that galaxies 
experience many mergers over the course of  more than 11 Gyr 
from $z\sim 3$ to the present day \citep{2009ApJ...702.1005S,
2009MNRAS.394.1956C,2011ApJ...742..103L,
2013A&A...553A..78L,2016MNRAS.458.2371R}. Most of these would 
be termed `minor mergers'' with $\Delta m_{\ast}/m_{\ast} < 1/4$. 
However, gas-rich (``wet") mergers can drive gas into the nucleus of 
galaxies, making the centers denser and more compact, gas-poor 
(``dry'') mergers produce remnants that are more extended than the 
progenitors \citep{2010MNRAS.401.1099H,2011MNRAS.415.3135C}. 
Several works have pointed out that a promising way to drive rapid 
size growth with relatively little growth in mass is for a galaxy to 
experience multiple dry, minor mergers \citep{2009ApJ...697.1290B,
2009ApJ...699L.178N,2012ApJ...744...63O,2013MNRAS.429.2924H,
2014MNRAS.444..942P}.

To summarize, the suggested physical sequence (as described by 
\cite{2013ApJ...765..104B}) is the following: (1) star-forming, extended 
galaxies are gas rich, and undergo dissipative processes to form 
compact SFGs. (2) This compaction is followed by rapid quenching of 
star formation, to form compact quiescent galaxies. Finally, (3) the 
galaxy sizes are gradually expanded via multiple, minor dry mergers. 
Although this picture is plausible and appealing, it is still an open 
question whether fully self-consistent numerical cosmological 
hydrodynamical simulations can qualitatively or quantitatively 
reproduce the observed evolution.

Recent cosmological hydrodynamic simulations have demonstrated 
that AGN feedback can reduce galaxy stellar mass by preventing 
cooling flows \citep[cf.][]{Fabian_1994} and quenching subsequent 
star formation in group and cluster sized halos \citep[see reviews 
by][for references]{2015ARA&amp;A..53...51S,2017ARA&amp;A..55...59N}.
Several recent studies have examined predictions for the evolution of 
the size-mass relations for both star forming and quiescent galaxies in 
hydrodynamic simulations of large cosmological volumes that include 
various implementations of AGN feedback
\citep{2014MNRAS.445..175G,2015MNRAS.449..361W,
2015MNRAS.450.1937C,2017MNRAS.465..722F,2018MNRAS.474.3976G}. 

While these studies found qualitative agreement with the observations, 
predictions for galaxy size and structure are known to be extremely 
sensitive to the details of how processes such as star formation, stellar 
feedback, and AGN feedback are implemented on sub-resolution scales 
(``sub-grid" physics) \citep[see discussion and references in][]
{2015ARA&amp;A..53...51S}. For example, the predicted size-mass 
relation for star forming galaxies is quite different in the original Illustris 
simulation and the revised Illustris-TNG simulation 
\citep{2018MNRAS.474.3976G}, but the physical reasons for this are 
not fully understood (S. Genel, private communication). 

It has been shown that simulations that include AGN feedback produce 
considerably better agreement with the observed size-mass relation for 
massive, quiescent galaxies, while simulations that omit AGN feedback 
produce massive galaxies that are too compact 
\citep{2013MNRAS.433.3297D,2015MNRAS.450.1937C,
2017ApJ...844...31C}. However, once again the physical processes at 
play are not fully understood. AGN feedback clearly reduces the gas 
fraction in early massive galaxies, leading to more dry merging, which, 
as discussed above, can drive rapid size growth. Furthermore, AGN 
feedback shuts off the late in-situ star formation that tends to rebuild a 
compact stellar core. 

In addition, a potentially important dynamical 
impact of AGN feedback on the dark matter and stellar 
distributions has been discussed recently. 
\citet{2012MNRAS.422.3081M,2013MNRAS.432.1947M} found that 
AGN feedback can produce a flat stellar and dark matter density core 
in a galaxy cluster scale zoom-in simulation, possibly via a coupling of 
several mechanisms such as orbital energy transfer by black holes 
through dynamical friction, and `central revirialization' 
\citep{2012MNRAS.421.3464P} after AGN-driven gas outflows. 

In this study, we make use of the high resolution simulations of massive 
galaxies evolving in a full $\Lambda$CDM context presented by 
\citet{2017ApJ...844...31C} to study the evolution of size, stellar mass, 
star formation rate, and compactness of galaxies during the critical 
transition phase $0<z<3$. These simulations utilize the ``zoom-in'' 
technique to attain higher mass and spatial resolution than the large 
cosmological volumes.  Furthermore, these simulations 
include a more detailed and physically motivated treatment of the 
feedback from radiatively efficient accretion onto supermassive black 
holes \citep[see Section 2.5 in][]{2017ApJ...844...31C} than previous 
cosmological simulations that have studied galaxy 
structural evolution. Our feedback model is motivated by observations 
of powerful winds seen in broad absorption line (BAL) AGN 
\citep{2009ApJ...706..525M,2009ApJ...692..758G,
2012ApJ...753...75C,2013MNRAS.436.3286A}. We include 
\emph{mechanical feedback} from the BAL winds, as well as 
\emph{radiative feedback} via Compton and 
photoionization heating \citep{2012ApJ...754..125C,
2015MNRAS.449.4105C}. Furthermore, we analyze a matched suite of 
simulations with and without AGN feedback, and delve in more detail 
into the physical processes responsible for 
driving the predicted structural evolution.

We describe our simulations in \S\ref{simulations}, present our results 
in \S\ref{result}, and conclude in \S\ref{discussion}.

\section{Simulations}\label{simulations}
We performed two sets of high-resolution, cosmological zoom-in
hydrodynamic simulations of 30 massive halos, with present-day
total masses of $1.4 \times 10^{12} \Msun$ $ \le M_{\mathrm{vir}} \le 
2.3 \times 10^{13} \Msun$ and present-day stellar masses of 
$8.2 \times 10^{10} \Msun \le M_{\ast} \le 1.5 \times 10^{12} \Msun$ 
(for the central galaxies). The physics implemented in our simulation
models includes star formation, supernova feedback, wind feedback 
from massive stars, AGB stars and metal cooling and diffusion. Our 
simulation code also incorporates a new treatment of mechanical and 
radiative AGN feedback which is implemented in a self-consistent way, 
launching high-velocity mass outflows. The simulation suite used in 
this study is presented in \cite{2017ApJ...844...31C}, and we refer the 
reader to that paper for further details. In the following we briefly 
summarize our simulations.

\subsection{The hydrodynamic simulation code: SPHGal}
We use SPHGal \citep{2014MNRAS.443.1173H}, a modified version of the 
parallel smoothed particle hydrodynamics (SPH) code GADGET-3 
\citep{2005MNRAS.364.1105S}. This code incorporates a density-independent 
pressure-entropy SPH formulation \citep{2001MNRAS.323..743R,
2013ApJ...768...44S,2013MNRAS.428.2840H}, to overcome the numerical 
fluid-mixing problems of classical SPH codes 
\citep[e.g.][]{2007MNRAS.380..963A}. It also includes an improved 
artificial viscosity implementation \citep{2010MNRAS.408..669C}, an 
artificial thermal conductivity \cite{2012MNRAS.422.3037R}, and a 
Wendland $C^4$ kernel with 200 neighboring particles 
\citep{2012MNRAS.425.1068D}. In order to ensure a proper treatment of
shock propagation and feedback distribution, we employ a time-step 
limiter that makes neighboring particles have similar time-steps 
\citep{2009ApJ...697L..99S,2012MNRAS.419..465D}. 

\subsection{Star formation, chemical enrichment and 
             stellar feedback model}
Following \cite{2013MNRAS.434.3142A}, stars are stochastically formed 
within over-dense regions when the gas density exceeds a density
threshold for the Jeans gravitational instability of the enclosed mass.
The star formation rate is computed as $d \rho_{\ast} /dt = \eta 
\rho_{\rm gas} /t_{\rm dyn}$ where $\rho_{\ast}$, $\rho_{\rm gas}$ and 
$t_{\rm dyn}$ are the stellar and gas densities, and local dynamical time
of gas particle, and we set the star formation efficiency as $\eta=0.025$.

The evolution of each star particle contributes to the chemical enrichment
of the interstellar medium (ISM) in our simulation during various mass
loss events. We allow chemical enrichment via winds driven by Type~I
Supernovae (SNe), Type~II SNe and asymptotic giant branch (AGB) stars with
the chemical yields adopted from \citet{1999ApJS..125..439I},
\citet{1995ApJS..101..181W}, and \citet{2010MNRAS.403.1413K}. We 
explicitly trace the mass in 11 chemical elements (H, He, C, N, O, Ne, Mg, 
Si, S, Ca and Fe), both for star and gas particles. We also allow the 
metal enriched gas particles to mix their metals with neighboring metal-poor 
gas particles via turbulent diffusion, using the standard SPH neighbor 
searches following \citet{2013MNRAS.434.3142A}. The net cooling rate is 
calculated based on individual chemical abundances, temperatures and 
densities of gas particles following \citet{2009MNRAS.393...99W}, 
accounting for a redshift dependent metagalactic UV/X-ray and cosmic 
microwave background with a modified \citet{2001cghr.confE..64H} spectrum.

The stellar feedback model is adopted from \citet{2017ApJ...836..204N},
and includes UV heating within \strom spheres around young massive stars,
three-phase SN feedback by both type I and type II SNe, and winds from 
dying low-mass AGB stars. Each of the mass loss events of the stellar 
particles explicitly contribute mass, metals, momentum and energy to the 
surrounding gas. First, the young star
particles gradually heat the neighboring gas to $T=10^4$ K within their HII
region limit \citep[\strom spheres][]{1939ApJ....89..526S} before they 
explode as SNe. 

We assume that a single SN event ejects mass in an 
outflow with a velocity $v_{\rm SN}=4,500 \kms$, a typical velocity of 
outflowing SN ejecta \citep{2012ARNPS..62..407J}. Depending on the physical
distance from the SN, each adjacent gas particle is affected by one of the
three successive SN phases: (i) momentum-conserving free expansion phase, 
(ii) energy-conserving Sedov-Taylor phase where SN energy is transferred
with 30\% as kinetic and 70\% as thermal, and (iii) the snowplow phase
where radiative cooling becomes important. In this `Snowplow' SN feedback
model, each SN remnant launches standard Sedov-Taylor blast-waves carrying
energy as 30\% kinetic and 70\% thermal, and both dissipate with distance
from the SN in its final pressure-driven snowplow phase.
In addition, the old stellar particles still contribute their mass, energy
and metal output via slow winds during an AGB phase. The energy and 
metal-enriched mass output from the old star particles  are transferred to
the adjacent gas particles in momentum-conserving manner. 

We assume a \citet{2001MNRAS.322..231K} initial mass function (IMF), and 
we note that over 30\% of the total mass in stars will be continuously ejected 
into the ISM via winds from SNe and AGB stars within $\sim$13 Gyr stellar
evolution. This metal-enriched mass output from stars not only fuels 
late star formation but also feeds the central super massive black
hole, inducing AGN activity \citep[see also][]{2010ApJ...717..708C}.

\subsection{Black hole formation, accretion, and AGN feedback model}
In the simulations,  new black hole seeds of mass $10^5 \Msunh$ are placed
in the center of galaxies when they initially reach a dark matter halo mass
of $10^{11} \Msunh$. The black hole seed mass and dark matter halo 
threshold mass are chosen to approximately follow the 
\citet{1998AJ....115.2285M} relation and the theoretical calculations of
black hole seed formation \citep[e.g.][]{2010A&amp;ARv..18..279V,
2017MNRAS.467.4180S}. We note that black hole seed mass makes a negligible 
contribution to the final black hole mass of the central galaxies in our
simulations.

The black hole can grow by merging with other black holes and by gas
accretion. When two black hole particles get closer than their local SPH
smoothing lengths and their relative velocities are less than the local
sound speed, they merge together. The black hole particle also grows in 
mass via accretion of surrounding gas with a Bondi-Hoyle-Lyttleton
accretion rate \citep{1939PCPS...34..405H,1944MNRAS.104..273B,
1952MNRAS.112..195B}, $\dot{M}_{\rm{inf}}= 
(4 \pi  G^{2} M_{\rm BH}^{2} \rho )/((c_{\rm s}^2+ v^{2})^{3/2})$, 
where $\rho$, $c_{\rm s}$, and $v$ are the density, the sound speed and 
the velocity of the gas relative to the black hole respectively. In order
to prevent the unphysical accretion of unbound gas outside the Bondi 
radius, we incorporate the soft Bondi criterion first introduced in
\cite{2012ApJ...754..125C}. We statistically limit the accretion of the
gas by the volume fraction of the gas particle lying within the  Bondi
radius, for example, the full accretion of the particle is allowed only 
when the total volume of a {\it smoothed} gas particle is contained within
the Bondi radius. 

The AGN feedback model we use is adopted from \citet{2012ApJ...754..125C,
2014MNRAS.442..440C} and consists of two main components: (1) mechanical 
feedback via {\it winds}, which carry energy, mass
and momentum into the neighboring gas, and (2) radiative feedback via 
Compton and photoionization heating from the X-ray radiation from the
accreting black hole. We also incorporate the radiation pressure 
associated with the X-ray heating, and the Eddington force. The emergent
AGN spectrum and metal line heating rate are taken from 
\cite{2004MNRAS.347..144S}. The winds are driven by radiation pressure on 
gas and dust on scales below those that we can simulate explicitly; we 
therefore treat them using a sub-grid model as outlined below. 

For the mechanical AGN feedback, the gas mass inflowing to the central
region contributes to the accretion onto the black hole and AGN-driven
winds. AGN winds carry mass with the kinetic energy rate
given as $E_{\rm outf,AGN} =0.5 \Mdotoutf v_{\rm outf,AGN}^2$,
where the AGN outflowing wind velocity is assumed to be 
$v_{\rm outf,AGN}=$10,000 $\kms$, motivated by observations of BAL 
winds \citep[e.g.][]{2013MNRAS.436.3286A}. This kinetic energy 
rate is proportional to the gas mass accreted onto the black hole as 
$\dot{E}_{\rm outf,AGN} \equiv  \epsw \Mdotacc c^2$, where $\epsw$ indicates 
the AGN feedback efficiency and is set to be $\epsw=0.005$, similar to the 
values used in other AGN feedback model implementations 
\cite[e.g.][]{2005Natur.433..604D}.

Based on our selection of the feedback efficiency $\epsw$ and the wind
velocity $v_{\rm outf,AGN}$, a fraction of the gas particles entering 
the accretion region is stochastically selected as wind particles. The
selected wind particles are kicked in a direction parallel or 
anti-parallel to the direction of angular momentum of each gas particle;
therefore the wind tends to be oriented perpendicular to the disk plane 
\citep{2004ApJ...616..688P}, when the black holes are surrounded by 
a rotating gas disk. 
The ejected wind particle shares its energy and momentum with two 
adjacent gas particles and produces a shock heated momentum-driven
flow with a ratio of kinetic to thermal energy similar to that in the 
standard Sedov-Taylor blast wave.

We also incorporate the heating via hard X-ray radiation from the accreting 
black hole following \cite{2005MNRAS.358..168S}. At the position of each 
gas particle, we calculate the net luminosity flux from all black holes in the 
simulated zoom-in area. The calculated flux is then converted to the net 
volume heating rate $\dot{E}$ via Compton and photoionization heating using 
\cite{2005MNRAS.358..168S} formulae. The radiation pressure from the X-ray
flux is also included, as every gas particle absorbing energy $\Delta E$
from X-ray radiation is given an additional momentum $\Delta p = \Delta E /c$
directed away from the black hole.

Finally, we include the Eddington force on electrons in the neighboring gas 
directed radially away from black holes, instead of artificially limiting the
gas accretion rate onto the black hole not to exceed the Eddington rate. 
In our simulations, we allow for super-Eddington gas accretion occasionally 
to occur. When this happens, the corresponding feedback effects naturally 
reduces the inflow and increases the outflow.

\cite{2015MNRAS.449.4105C} showed that the mechanical  AGN feedback 
via broad absorption line winds produces stronger effects on the galaxy 
compared to the thermal feedback treatment 
\citep[e.g.][]{2005MNRAS.361..776S}. The feedback energy injected via 
mechanical winds rather than in thermal form more effectively drives the 
gas out of the galaxy via galactic outflows \citep{2018ApJ...860...14B} and 
quenches the star formation for a longer timescale.

\subsection{Initial conditions}
We use the cosmological `zoom-in' initial conditions described in 
\citet{2010ApJ...725.2312O}. A sub-volume for the zoom-in initial condition 
is extracted from a larger volume dark matter only simulation using a flat 
cosmology with parameters obtained from WMAP3 \citep[][$h=0.72, \;
\Omega_{\mathrm{b}}=0.044, \; \Omega_{\mathrm{dm}}=0.216, \;
\Omega _{\Lambda}=0.74, \; \sigma_8=0.77 $, and 
$\mathrm{n_s}=0.95$]{2007ApJS..170..377S}. All dark matter particles 
close to the halos of interest are traced from redshift zero and then 
replaced with higher resolution gas and dark matter particles. 

We re-simulate the new high resolution initial conditions from 
redshift $z=43$ to $z=0$. The simulation sets used in this study have 
mass resolution for the baryonic particles (star and gas) of 
$m_{*,gas}=5.8 \times 10^{6} \Msun$, and the dark matter particles of 
$m_{\mathrm{dm}} = 3.5 \times 10^{7} \Msun$. We adopt the co-moving 
gravitational softening lengths $\eps_{\mathrm{gas,star}} = 400 \rm \pch $ 
for the baryonic particles and $\eps_{\mathrm{halo}} = 890 \rm \pch$ for 
the dark matter. The adopted resolution is summarized in Table~\ref{tab:resolution}.

\begin{table}
   \begin{center}
   \caption{Summary of the simulation resolution}
    {
   \begin{tabular}{c|c|c}\hline\hline \label{tab:resolution}
Mass Ratio &  softening length $\eps$ & particle mass  \\
 & $\rm \pch$& $\Msun$ \\
  \hline
 dark matter & 890 & $3.5 \times 10^{7}$ \cr
 baryon& 400 & $5.8 \times 10^{6}$ \cr
  \hline\hline
   \end{tabular}}
   \end{center}
 \end{table}
 
The simulated halo masses range from  $1.4 \times 10^{12} \Msun$ 
$ \lesssim M_{\mathrm{vir}} \lesssim 2.3 \times 10^{13} \Msun$ at $z=0$ 
and the stellar masses of central galaxies are $8.2 \times 10^{10} \Msun 
\lesssim M_{\ast} \lesssim 1.5 \times 10^{12} \Msun$ at present day. 
These galaxies are well resolved with $\approx 2.5 \times 10^4 - 4.8 
\times 10^5$  stellar particles within the virial radius 
($R_{200}$, the radius where the spherical 
over-density drops below 200 times the critical density of the universe at
a given redshift). The halo virial mass $M_{200}$ is the mass contained 
within a sphere with radius $R_{200}$.

In order to study the effects of AGN feedback on the star 
formation, size and compactness of the galaxies, we run the full
set of simulations with two different sets of physics:

{\bf (1) NoAGN:} No black holes and no AGN feedback. This model is 
comparable to the results of \cite{2010ApJ...725.2312O,
2012ApJ...744...63O}, but note that we use an alternative 
density-independent formulation of SPH designed to treat contact 
discontinuities more accurately, and include an artificial viscosity and
an energy diffusion implementation which results in more cooling. We also
include metal enrichment and metal line cooling which also enhances the
cooling process. Moreover, we adopt a different recipe for stellar feedback 
\citep[see][for a further exploration of the effects of these changes]{2017ApJ...844...31C}.

{\bf (2) WithAGN:} all of the same physical models as in the NoAGN
model, with the addition of black holes and mechanical and radiative AGN 
feedback.

In this study, we only consider the central galaxies within the simulated
halos. The stellar mass and the star formation rate of the galaxy are measured 
within 10 percent of the virial radius $r_{10}=0.1 \times R_{200}$. Then we 
determine the effective radius of the galaxy $\reff$ by determining the 
mean values of the half-mass radii of stars within $r_{10}$ projected 
along 20 randomly chosen directions of the main stellar body.

\section{Results}\label{result}
In \cite{2017ApJ...844...31C}, we showed that AGN feedback, as 
implemented in our simulations, can change the fate of massive 
galaxies, effectively quenching late star formation, turning the majority 
of massive galaxies into quiescent galaxies, and reducing their final 
stellar mass by almost a factor of four. AGN feedback effectively reduces 
the in-situ star formation in the central galaxy where the most massive 
black holes are located. Our simulated galaxies without AGN show much 
higher rates of in-situ star formation in the innermost regions of the 
galaxies, as the gas reservoir is constantly refilled by ``recycled gas'' from 
SN and AGB winds from the old stellar population. In this section we 
show that AGN feedback also affects various structural properties of 
galaxies, including galaxy sizes, stellar core density, and compactness.

\subsection{Evolution in the size-mass plane}
\begin{figure*}
  \includegraphics[width=\textwidth]{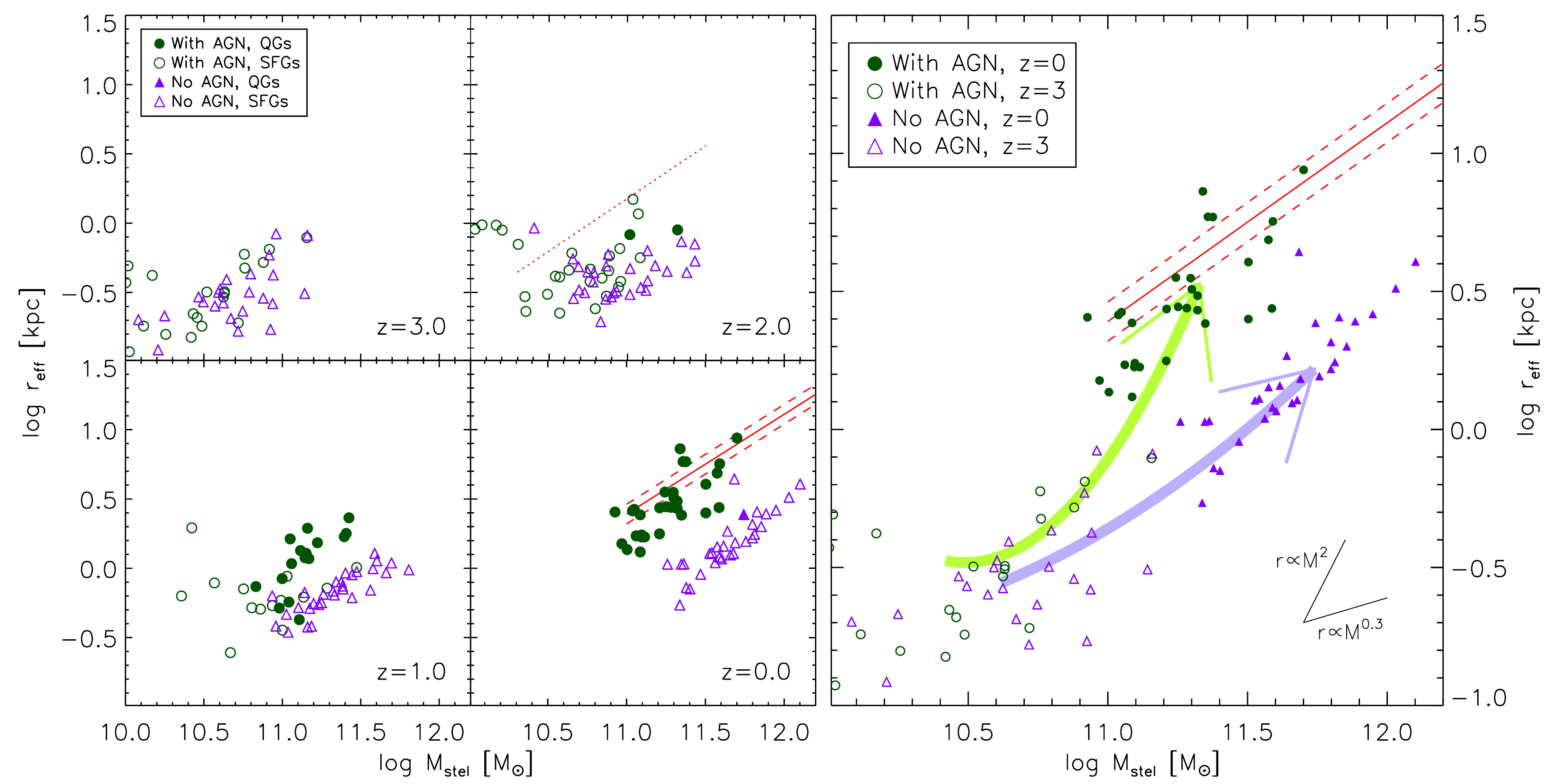}
    \caption{Size evolution of simulated galaxies with AGN feedback (green
    circles) and without AGN feedback (purple triangles) from $z=3$ to $z=0$.
    The observed size-mass relation of present-day quiescent galaxies 
    from \cite{2009ApJ...706L..86N} is shown in red solid line (with 
    1$\sigma$ scatter in red dotted lines). The size-mass relation for 
    high-redshift galaxies ($z=2.25$) from CANDELS 
    \citep{2014ApJ...788...28V} is shown in red dotted line. {\it (left)}
    projected stellar half-mass radii vs. stellar mass for simulated 
    galaxies  are shown respectively at $z=3,2,1$ and $0$. Quiescent 
    galaxies with low specific star formation rates (sSFR 
    $ \le 0.3 t_{\rm H}$) are shown in solid symbols, while star forming 
    galaxies (sSFR $ \ge 0.3 t_{\rm H}$) are shown in open symbols. 
    Galaxies simulated with AGN have their star formation quenched starting 
    at $z \sim 2$, and then start to evolve on a steeper track on the 
    size-mass relation. {\it (right)} Average tracks of simulated 
    galaxies on the size-mass plane from $z=3$ to $z=0$ are shown for 
    two models, with AGN feedback in green arrow, and without AGN feedback
    in purple arrow respectively. The points show the size-mass relation 
    of the simulated galaxies at $z=3$ in open symbols and at $z=0$ in 
    filled symbols respectively.       
 \label{fig:r_vs_mstar}}
\end{figure*}

\begin{figure}
 \includegraphics[width=\columnwidth]{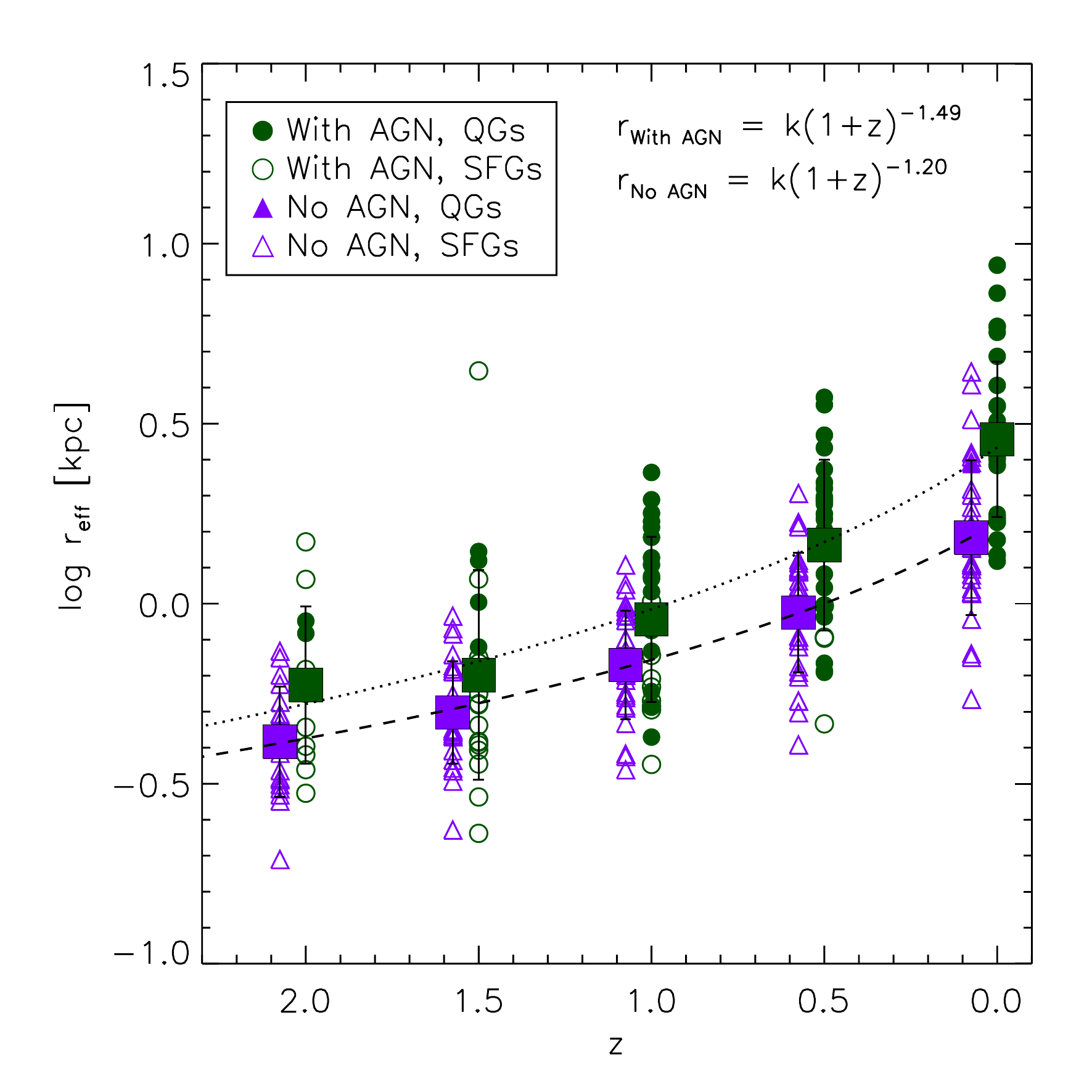}
 \caption{The redshift evolution of projected stellar half-mass radii of 
  galaxies with $\Mstar > 6.3 \times 10^{10} \Msun$. The galaxies 
  simulated with AGN are shown in solid green (quiescent) and in open green 
  (star forming) circles, and the galaxies simulated without AGN feedback
  are shown in solid purple (quiescent) and open purple (star forming) triangles. 
  The and purple squares show the mean sizes at a given redshift for 
  with AGN and no AGN models respectively. NoAGN model galaxies are offset 
  by 0.1 in redshift for clarity. The black lines show the result of 
  a power-law fit for the NoAGN (dashed) and withAGN (dotted) simulations.
 \label{fig:r_vs_z} }
\end{figure}

\begin{figure}
 \includegraphics[width=\columnwidth]{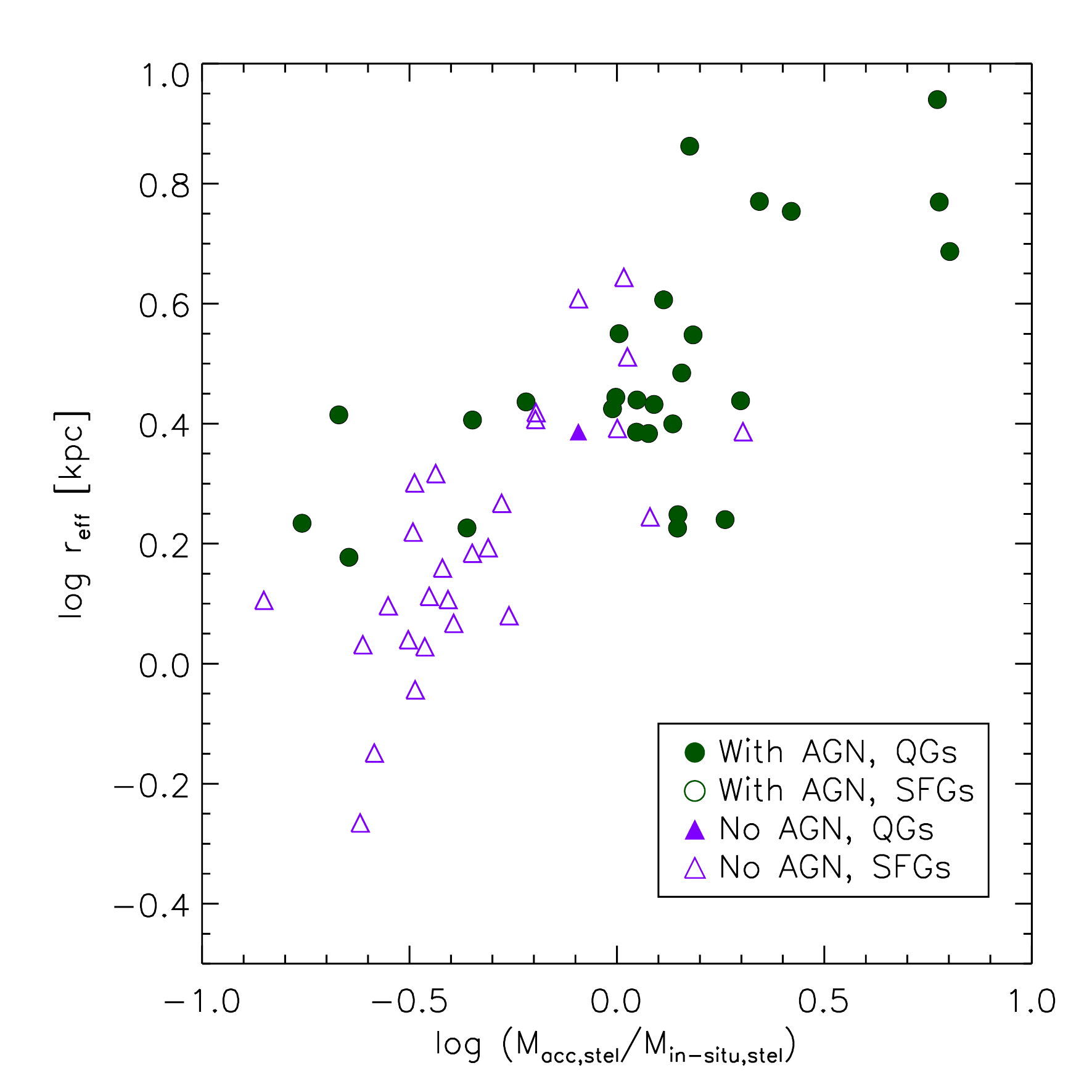}
 \caption{Projected stellar half-mass radii at $z=0$ as a function of
 the fraction of accreted stellar mass to the stellar mass formed in-situ 
 for the WithAGN model (green circles) and NoAGN model (purple 
 triangles). Quiescent galaxies with low specific star formation rates (sSFR 
 $ \le 0.3 t_{\rm H}$) are shown in solid symbols, and star forming 
 galaxies (sSFR $ \ge 0.3 t_{\rm H}$) are shown in open symbols. 
 For both WithAGN and NoAGN models,
 we find a correlation between galaxy size and the fraction of accreted 
 stars, i.e., galaxies with a higher fraction of accreted stars have 
 larger sizes. This suggests that the accretion of stellar mass through 
 mergers is a main driver of  the size growth of massive galaxies.
 \label{fig:r_vs_fraction} }
\end{figure}


\begin{figure*}
 \includegraphics[width=\textwidth]{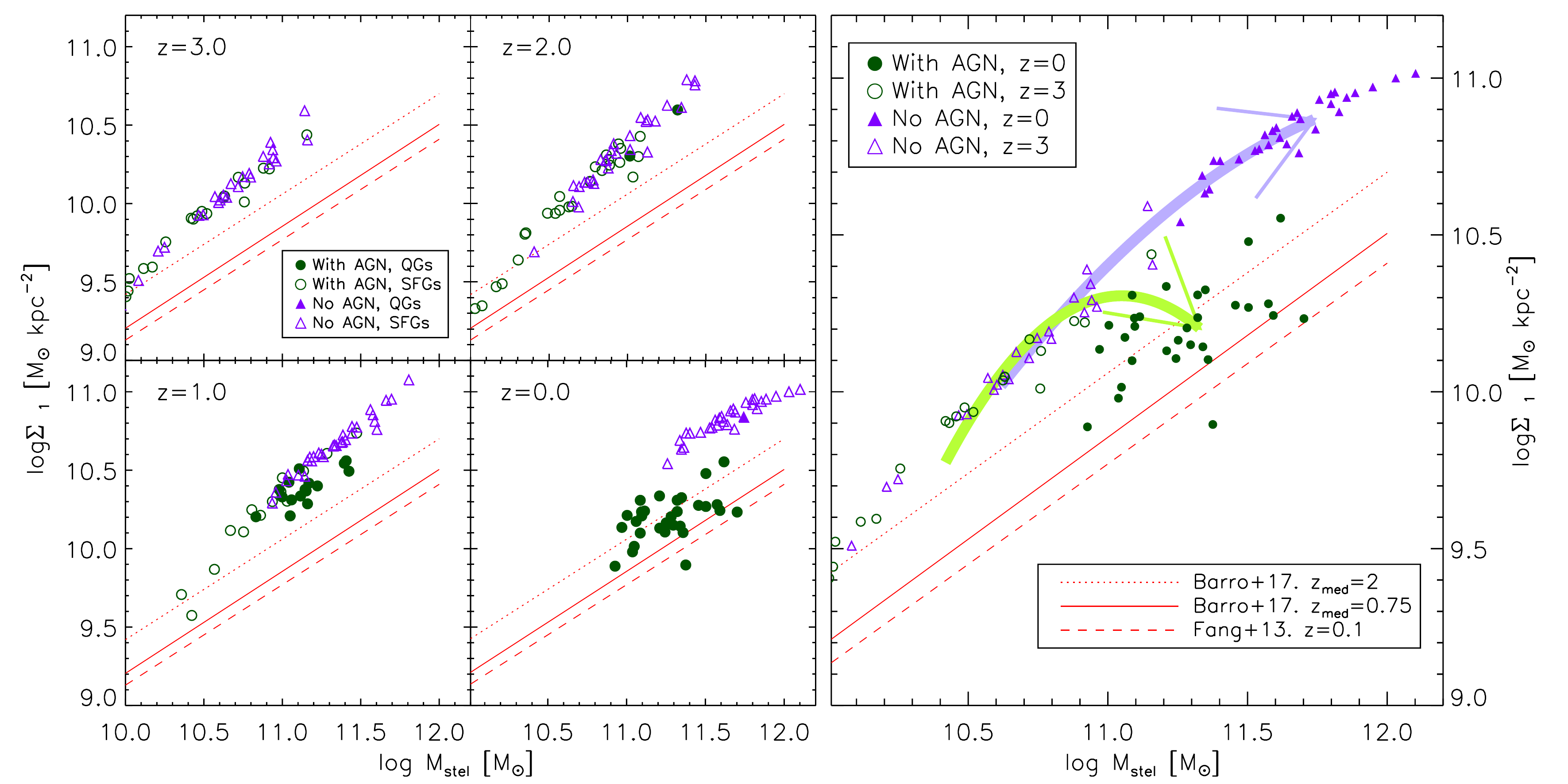}
  \caption{Projected stellar core density, $\Sigma_{1} =
   M_{\ast} (< 1 \kpc) / \pi (1 \kpc)^2 $ vs. stellar mass of 
   simulated galaxies with AGN feedback (green circles) 
   and without AGN feedback (purple triangles) from $z=3$ to $z=0$. The 
   observed $\Sigma_{1}$-$\Mstar$ relation of quiescent galaxies  
   is shown for $1.4 <z<2.2$ (red dotted line), for $0.5<z<1.0$ (red solid line)
   from  \citet{2017ApJ...840...47B}, and for $z \sim 0.1$ (red dashed line) 
   from  \cite{2013ApJ...776...63F} respectively.
   {\it (left)} projected stellar core density within 1 kpc vs. stellar 
   mass for simulated galaxies are shown respectively at $z=3,2,1$ and
   $0$. Quiescent galaxies with low specific star formation rates (sSFR 
   $ \le 0.3 t_{\rm H}$) are shown in solid symbols, while star forming 
   galaxies (sSFR $ \ge 0.3 t_{\rm H}$) are shown in open symbols. 
   Quenched galaxies in AGN feedback model decrease their core density 
   with time since $z \sim 2$, while galaxies simulated without AGN keep
   increasing their core density. {\it (right)} Average tracks of 
   simulated galaxies on the $\Sigma_{1}$-$\Mstar$ plane from $z=3$ to 
   $z=0$ are shown for two models, with AGN feedback in green, and 
   without AGN feedback in purple respectively. The points show the 
   core density-mass relation of the simulated galaxies at $z=3$ in 
   open symbols and at $z=0$ in filled symbols respectively. 
   \label{fig:sigma1_vs_mstar}}
\end{figure*}

In left panels of Figure~\ref{fig:r_vs_mstar}, we show the evolution of
the galaxies simulated with and without AGN feedback in the projected 
half-mass radius (``size'') vs. stellar mass plane from redshift $z=3$ to $z=0$. 
We show the sizes and stellar masses of the 30 central galaxies at $z=0$, 
and their most massive progenitors at $z=1,2$ and $3$.  In order to exclude 
outliers caused by galaxy interactions (e.g. ongoing mergers), we use the 
median value of the sizes that we obtained for three snapshots around the 
target redshift. We divide our galaxies into star forming (open symbols) and 
quiescent galaxies (solid symbols) based on the commonly adopted dividing 
line at sSFR$=0.3/t_{\rm H}$, where $t_{\rm H}$ is the age of the universe 
at each redshift \citep[e.g.][]{2008ApJ...688..770F}.  We adopt this method
as this divider based on sSFR is roughly equivalent to the commonly adopted 
method that separates quiescent and star forming  galaxies in 
color-color space up to $z \sim 2.5$ as described in \cite{2009ApJ...691.1879W,
2010ApJ...713..738W,2013ApJ...777...18M}.
Note that we only consider the central galaxies in the simulated halos. 
The central galaxies in group environments may experience accelerated size 
growth compared to galaxies in the field, via minor mergers which are more
common in high density environments
  \citep[see][]{2013ApJ...778L...2C,2018ApJ...856....8C}.

At $z=3$, all progenitor galaxies simulated with and without AGN feedback
are very compact ($\reff < 1$ kpc), and the two models do not show a 
significant difference in size while NoAGN galaxies are slightly more massive 
than WithAGN galaxies. At $z=2$, both WithAGN and NoAGN galaxies are 
still compact, compared to the size-mass relation for high-redshift galaxies 
($z=2.25$) from CANDELS \citep{2014ApJ...788...28V}. However, by $z=1$, 
the {\it quenched} galaxies among those simulated with AGN feedback start 
to diverge and form a size-mass relation clearly separated from that of 
NoAGN feedback galaxies. By $z=0$, all galaxies are quenched in the 
simulations with AGN feedback, and they closely follow the observed 
size-mass relation of the local quiescent galaxies from SLACS sample
\citep{2009ApJ...706L..86N}. 

On the other hand, NoAGN galaxies show much less size growth since 
$z\sim2$, evolving with a much shallower slope in the size-mass plane. The 
present-day sizes of galaxies simulated without AGN feedback are $\sim 5$ 
times smaller compared to the observed relation at a given stellar mass, 
consistent with \citet{2015MNRAS.450.1937C}. In the absence of AGN 
feedback, we have continuous star formation in the central regions of 
galaxies, which results in a concentrated stellar mass profile as discussed in 
\citet{2012MNRAS.422.3081M}. In addition, the ``outer'' size growth via
addition of stars from minor dry mergers is less significant in the NoAGN 
simulations, as in situ star formation always dominates over the accreted 
star component (see Figure 9 in \citet{2017ApJ...844...31C}, and also 
\citet{2012MNRAS.425..641L,2013MNRAS.433.3297D,
2013MNRAS.436.2929H,2016MNRAS.463.3948D}.

Since $z\sim2$, the galaxies without AGN feedback mainly accumulate their 
stellar mass via in-situ star formation (stars formed from gas within the main 
progenitor), while the galaxies with AGN feedback mainly grow their stellar 
mass by accreting stellar mass via minor mergers, after their in-situ star 
formation is quenched. This difference in the evolutionary paths of galaxies in 
the two models in the size-mass plane is illustrated in the right panel of 
Figure~\ref{fig:r_vs_mstar}, where we show the average tracks of simulated 
galaxies for the two models. The two models with and without AGN feedback 
follow very different evolutionary tracks. Until $z \sim 2$, galaxies in both 
models grow in mass and gradually increase their density, showing little 
growth in size. Then, after the in-situ star formation is quenched (as shown in 
left panels), the galaxies simulated with AGN feedback show much steeper 
evolution in the size-mass plane.

We note that the galaxies simulated with AGN feedback follow a very similar 
track to that suggested based on observations by \cite{2015ApJ...813...23V}, 
as discussed in the Introduction. Star-forming galaxies in our models evolve 
along the shallower track in the size-mass plane, $\Delta r\propto\Delta M^{0.3}$, 
until their star formation is quenched by the AGN-driven winds. After 
quenching, galaxies grow along a steeper track in the size-mass 
plane, with $\Delta r\propto\Delta M^{2}$. On the other hand, galaxies 
simulated without AGN feedback remain gas rich, and continue to 
evolve along the much shallower track with $\Delta r\propto\Delta M$.

Recently \citet{2018MNRAS.474.3976G} also found such a turnover in the 
size-mass plane in the evolution of massive galaxies in the 
{\it Illustris}-TNG simulation. They found that galaxies have this steep
size evolution as a function of the added stellar mass after they quench 
their star formation --- mainly induced by their black hole driven
kinetic and thermal  feedback  \citep{2017MNRAS.465.3291W} as in this 
study --- and the size growth of the most massive galaxies occurs mostly 
during their quiescent phase.
 
To quantify the size evolution of our simulated galaxies as a function of 
redshift, we fit a power law in $(1+z)^\alpha$ as is frequently done in 
observational studies \citep[e.g.][]{2010ApJ...717L.103N,
2011ApJ...739L..44D}. We show the size evolution of galaxies with 
$\Mstar > 6.3 \times 10^{10} \Msun$ from $z=2$ to the present day in 
Figure~\ref{fig:r_vs_z}. The median sizes of galaxies in each redshift bin 
are shown as squares and we show the power-law fit to the median sizes 
with black lines. Overall, galaxy sizes are smaller in the NoAGN model at 
all redshifts, and the difference becomes even larger by $z=0$ as WithAGN 
galaxies show more rapid size growth per added stellar mass during the 
quenched phase as shown in Figure~\ref{fig:r_vs_mstar}. Between redshift 2 
and 0, our simulated galaxies grow on average by a factor of 5. Our 
power-law fit to galaxy sizes simulated with AGN is $\alpha = -1.49$, which 
is in good agreement with observations \citep{2014ApJ...788...28V}. 

In Figure~\ref{fig:r_vs_fraction}, we show the projected stellar half-mass
radii at $z=0$ as a function of the fraction of stellar mass formed in other 
galaxies and accreted onto the main progenitor, to the stellar mass 
formed in-situ, for galaxies simulated with and without AGN feedback. 
There is a clear correlation between the sizes and the relative amount of 
accreted and in-situ formed stars for WithAGN as well as NoAGN feedback 
models. This indicates that the accretion of stellar mass drives the growth of 
the effective radius of massive galaxies for both models. For galaxies whose 
mass is dominated by the accreted stellar component $M_{\rm acc} > 
M_{\rm in-situ}$, the two models show similar sizes at fixed accreted star
fraction. However, for galaxies dominated by in-situ formed stars, 
$M_{\rm in-situ} > M_{\rm acc} $, galaxies simulated without AGN show 
smaller sizes at fixed fraction. This is partly because of the fact that 
completely dry mergers are relatively rare due to the high gas fractions 
in satellite galaxies without AGN feedback, while merging satellite galaxies
also have black holes and corresponding AGN feedback in the WithAGN 
feedback model. Moreover, galaxies with AGN are also more efficiently 
puffed up via gas outflows \citep{2008ApJ...689L.101F}. The slow expulsion 
of gas while the AGN is quiescent can induce adiabatic expansion of the stellar
component \citep[e.g.][]{1980ApJ...235..986H}, and the rapid mass loss 
driven by fast AGN winds can further elevate this puffing-up process 
\citep{1979ApJ...230L..33B}. We discuss and illustrate this puffing up 
process of the stellar system in the next subsection.

\begin{figure}
 \includegraphics[width=\columnwidth]{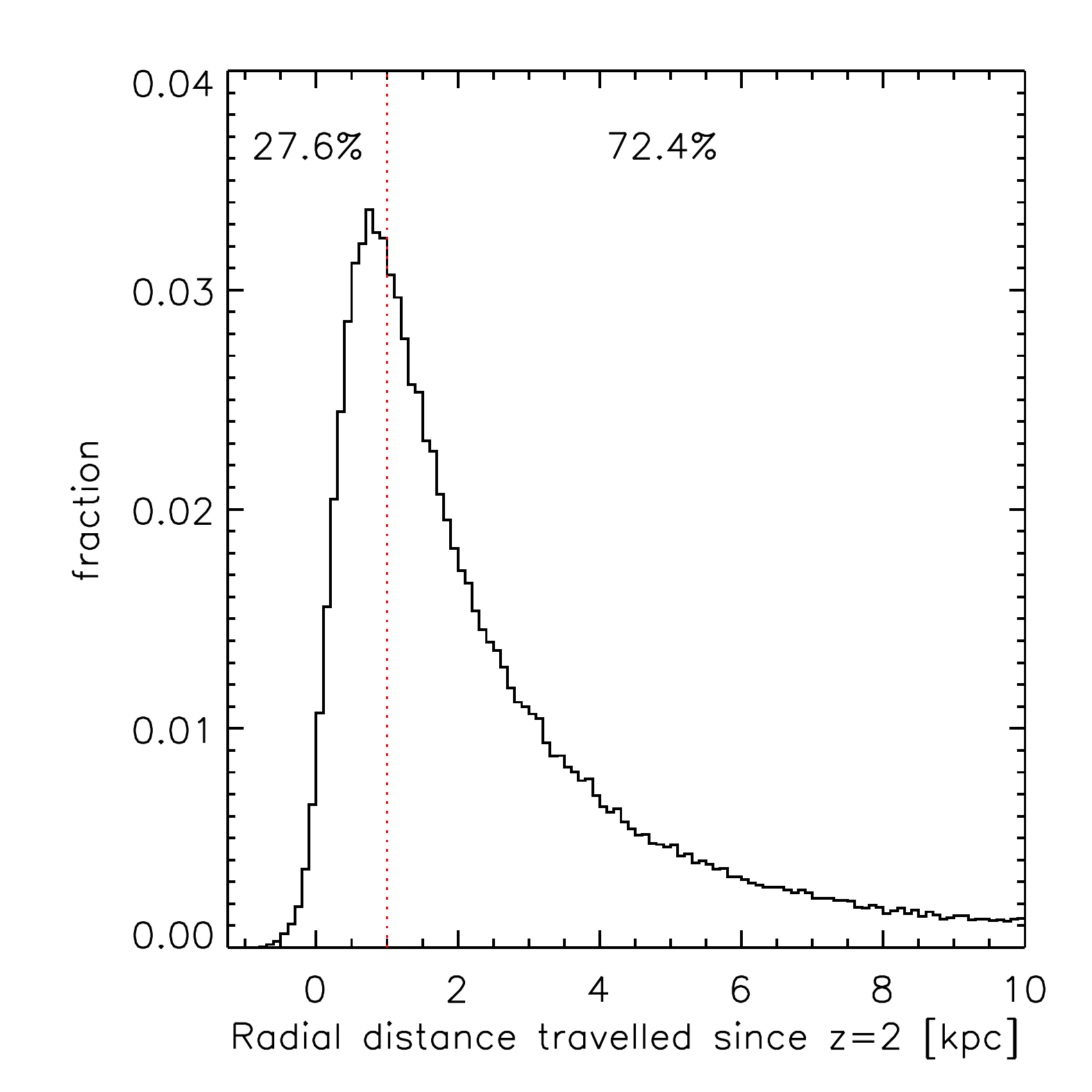}
 \caption{Stellar radial migration distance since $z=2$ measured for all
 star particles within the central region $R <1$ kpc for 30 galaxies 
 simulated with AGN.
 \label{fig:puff} }
\end{figure}

\begin{figure*}
 \includegraphics[width=\textwidth]{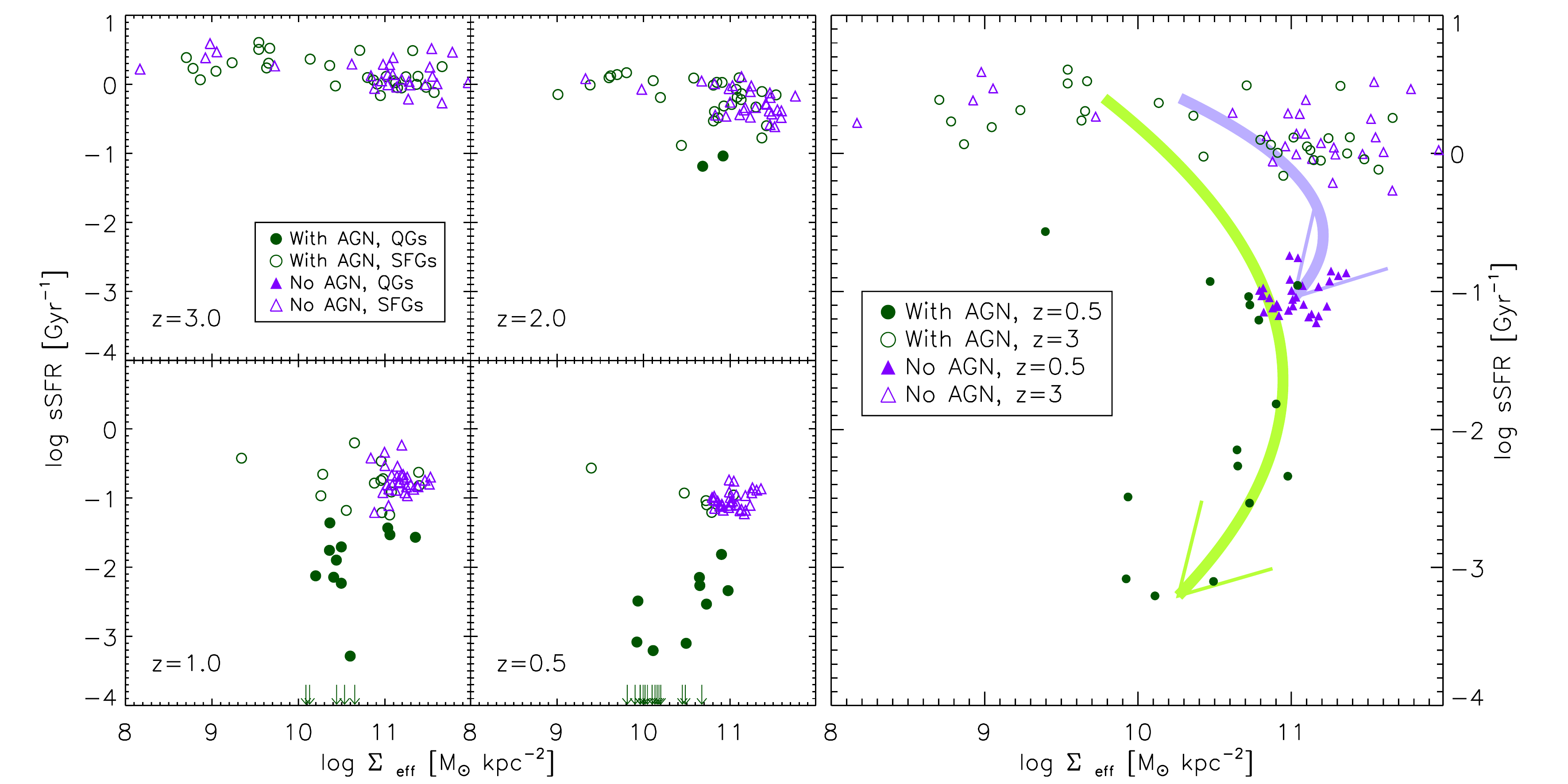}
 \caption{Evolution of the galaxies simulated with AGN feedback (green
 circles) and without AGN feedback (purple triangles) on the 
 sSFR-$\Sigma_{\rm eff} (= \Mstar /\pi \reff^{2})$ plane from $z=3$ to 
 $z=0.5$. {\it (left)} specific star formation rate  vs. projected stellar
 density measured within the effective radius  for simulated galaxies are
 shown respectively at $z=3,2,1$ and $0.5$. As in 
 Figure~\ref{fig:r_vs_mstar} and \ref{fig:sigma1_vs_mstar}, we show
 quiescent galaxies with low specific star formation rates (sSFR 
 $ \le 0.3 t_{\rm H}$) in solid symbols, and star forming galaxies (sSFR 
 $ \ge 0.3 t_{\rm H}$) in open symbols. We indicate the galaxies with
 specific star formation rates of $\rm log~sSFR < -4$  with downward arrows 
 showing their effective stellar densities. Galaxies simulated with AGN 
 feedback start to quench their star formation at $z\sim2$, and gradually
 move downwards on the sSFR-$\Sigma_{\rm eff}$ plane, and move toward the 
 left afterwards.  {\it (right)} Average tracks of simulated galaxies on 
 the sSFR-$\Sigma_{\rm eff}$  plane from $z=3$ to $z=0.5$ are shown for 
 two models, with AGN feedback in green, and without AGN feedback in
 purple respectively. The points show the location of the simulated 
 galaxies on the sSFR-$\Sigma_{\rm eff}$ at $z=3$ in open symbols and at 
 $z=0$ in filled symbols respectively. 
\label{fig:sSFR_sigma} }
\end{figure*}

\begin{figure*}
 \includegraphics[width=\textwidth]{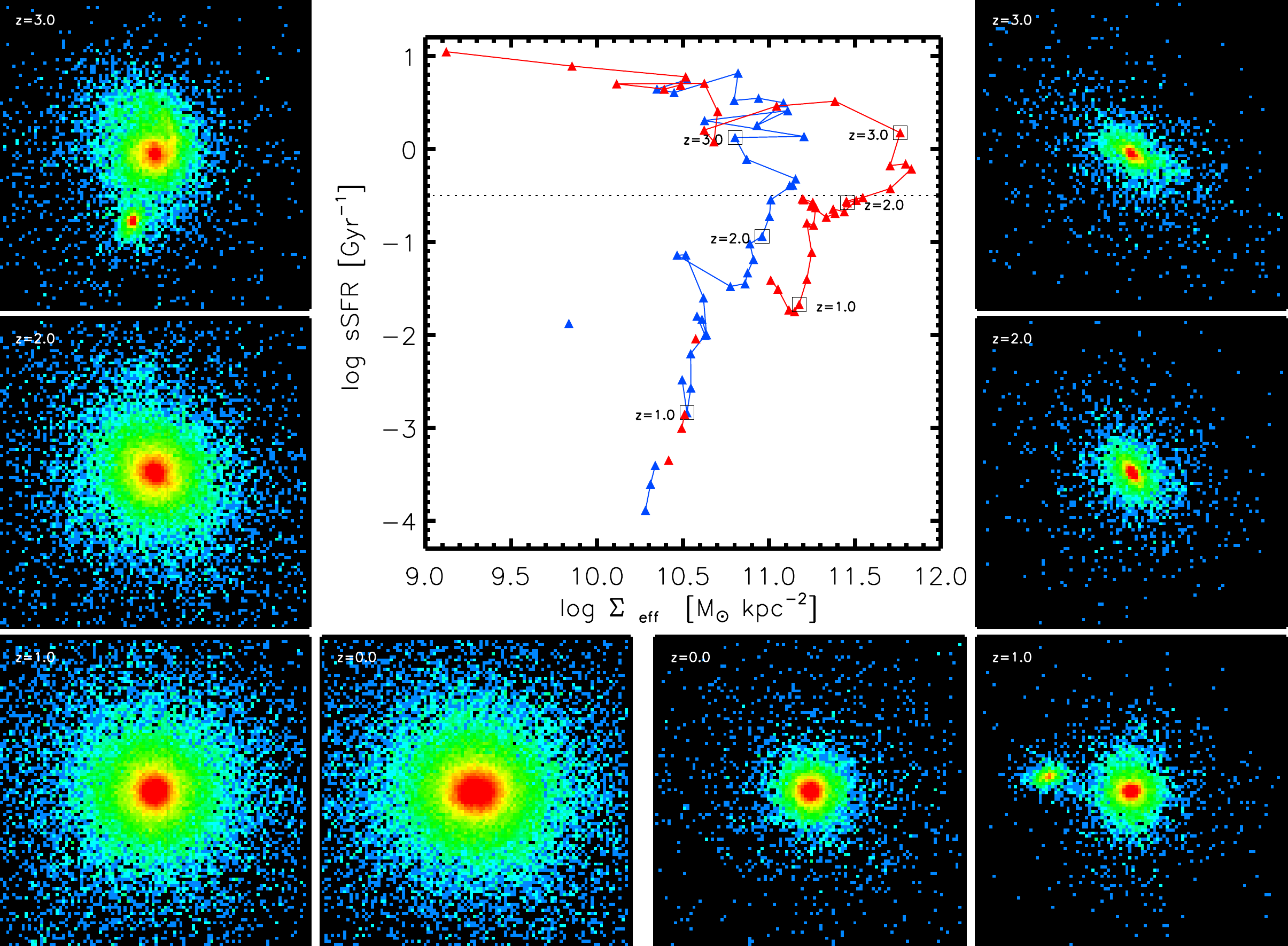}
 \caption{ {\bf Center panel:} Evolutionary tracks
 of two example galaxies on the plane of sSFR-$\Sigma_{\rm eff}$.
 The redshifts 
 $z=3,2,1$ are marked along the tracks by open squares. The fixed constant
 threshold $\rm log(sSFR)=-0.5$ adopted in \citet{2013ApJ...765..104B} is
 shown with black horizontal dotted line. {\bf Four left
 panels:} Projected stellar density maps at redshift $z=3,2,1$ and $z=0$ 
 of the example galaxy with $M_{\ast, z=0}=3.2 \times 10^{11} \Msun$ at 
 $z=0$ (shown in blue in the center panel). The snapshots are 100 $\kpc$ 
 on a side, and redder color indicates a higher density. {\bf Four right 
 panels:} Same as left panels but for the example galaxy with $M_{\ast, 
 z=0}=9.3 \times 10^{10} \Msun$ at $z=0$ (shown in red in the center 
 panel). 
\label{fig:track1}} 
\end{figure*}

\subsection{Central stellar density evolution}
\citet{2012ApJ...760..131C} proposed the quantity $\Sigma_{1}$, the 
projected stellar mass surface density within the central 1~kpc, as a 
robust measure of galaxies' core density, and showed that at 
$z\sim 0.5$--0.8, $\Sigma_{1}$ is more strongly correlated with 
quiescence than stellar mass or the effective surface density 
$\Sigma_{\rm eff}$. \citet{2013ApJ...776...63F} showed that 
$\Sigma_{1}$) shows a tight scaling relation with stellar mass for 
quiescent galaxies at $z\sim 0.1$, and that a stellar mass-dependent 
$\Sigma_{1}$ threshold is a better predictor of quiescence than is 
stellar mass alone. \cite{2017ApJ...840...47B} studied the structural 
properties of massive galaxies in the CANDELS/GOODS-S field and 
found that this tight scaling relation in $m_*$ vs. $\Sigma_{1}$ holds 
for quiescent galaxies out to $z\sim 2.5$, but the zero point of this 
relation has declined by $\sim 0.3$ dex since $z=2$. 
\citet{2014ApJ...791...45V} also found that the number density of 
galaxies with dense cores has decreased from $z=2.5$ to today. 
Both papers suggested that stellar mass loss and subsequent 
adiabatic expansion can explain this observed decrease in core 
density.

In order to study the evolution of the stellar core density of our 
simulated galaxies, we measure the projected central stellar mass 
surface density within $r<1 \kpc$, 
$\Sigma_{1} \equiv M_{\ast} (<\rm 1 \kpc) / \pi (1 \kpc)^2 $, and show
the evolution of all galaxies in the stellar mass versus $\Sigma_{1}$ plane 
from $z=3$ to $z=0$ in Figure~\ref{fig:sigma1_vs_mstar}. The core densities 
$\Sigma_{1}$ indicated here are the mean values of the surface stellar densities
projected along the three principal axes. We note that we only focus on the
differential effect as this quantity is affected by gravitational softening.

In left panels of Figure~\ref{fig:sigma1_vs_mstar} we show $\Sigma_{1}$ as 
a function of stellar mass for 30 central galaxies at $z=0$, and their most 
massive progenitors at $z=1,2,3$  as in Figure~\ref{fig:r_vs_mstar}. We also 
separately show the star forming and quiescent galaxies  (open and filled 
symbols respectively) to study the effect of star formation quenching on the 
evolution of the core stellar density. Until $z\sim2$, galaxies in both models 
show a tight correlation between $\Sigma_{1}$ vs. stellar mass and do not 
show significant differences in the relationship. By $z=1$, the quenched 
galaxies among those simulated with AGN feedback start to show 
{\it decreased} core densities and diverge from the previously established 
relation, while galaxies without AGN feedback continue to increase their core 
densities as well as stellar mass.  
 
All galaxies are quenched in the WithAGN model by the present day, and they form 
a clearly separated relation from that of the NoAGN galaxies. While WithAGN galaxies 
are in better agreement with the observed relation from \citet{2017ApJ...840...47B} and  
\cite{2013ApJ...776...63F}, NoAGN galaxies show $\sim 0.5$ dex 
higher core densities at a given stellar mass than observed. This high core density in 
NoAGN galaxies is attributed to the continuous in situ star formation of the galaxies, 
which is due to the lack of star formation quenching mechanisms in the central region. 
NoAGN galaxies have extended star formation until the present day, especially in the 
innermost region of the galaxies, as the gas reservoir in the galactic center is constantly 
refilled by ``recycled gas'' from SN and AGB winds in the old stellar population.
Although WithAGN galaxies show much lower core densities than NoAGN galaxies,
they are still 0.2-0.3 dex denser than the observations \citep{2013ApJ...776...63F} at low 
redshift. This is presumably due to overcooling at high redshift as galaxies in both 
models are already too dense in the cores at $z=2$ and 3.

In the right panel of Figure~\ref{fig:sigma1_vs_mstar}, we show the average tracks of 
simulated galaxies in the $\Sigma_{1}$-stellar mass plane for the WithAGN and NoAGN 
models. Until $z \sim 2$, galaxies in both models grow their stellar mass as well as 
core densities following a tight correlation. NoAGN galaxies keep following this 
evolutionary track until $z=0$ increasing their core densities, though the slope becomes
a bit shallower by the present day. The core densities of galaxies simulated with AGN feedback,
however, start to decline when the in-situ star formation is suppressed (as shown in left
panels). This transition seems to happen when galaxies reach $\Mstar \sim 10^{11} \Msun$ 
where a majority of galaxies start have their star formation quenched by AGN feedback.
Overall, the galaxies with AGN feedback show almost a 0.3 dex decrease in core 
density from redshift 1 to 0, consistent with recent observations by \citet{2017ApJ...840...47B}. 

We find that, in our simulations, the decrease in core density is a result of a combination 
of several physical processes. One contributor is stellar mass loss, i.e., old stars 
gradually lose their mass by stellar 
evolution via supernova and AGB winds. Without additional star formation, the central 
stellar density naturally decreases due to this effect. In our simulations, star 
particles gradually return mass to the adjacent gas particles via SN or AGB driven winds. 
In order to check how much mass stellar particles have lost due to stellar evolution, we trace the 
evolution of all central star particles within $r <$ 1 kpc at $z=2$ over time until $z=0$. 
On average, these central star particles have lost $\sim$ 10 percent of their mass since 
$z=2$ via stellar evolution. We note that with our assumed IMF, over 30\% of the total mass in 
newly formed stars will eventually be returned to the gas phase via winds over $\sim$13 Gyr of stellar 
evolution, but the bulk of the mass loss occurs when the stellar populations are very young,
via SN explosions \citep[see also][]{2011ApJ...734...48L}. 
Therefore, the relatively old stellar population already in place in the 
central region of the galaxies only shows moderate stellar mass loss since $z\sim2$, and 
this cannot fully explain the $\sim 50$ decrease in central stellar densities in WithAGN 
galaxies.

Secondly, the expansion of the collisionless particles in the central region after gas 
ejection by outflows can `puff up' the central region due to a decrease in the gravitational potential, leading 
to a more diffuse stellar core. The effect of outflows on the central stellar density depends on
the amount of mass ejected as well as on the timescale of ejection. We have two 
channels for this `puffing up' process, depending on the timescale of the mass loss: an 
adiabatic expansion after slow gas mass loss, and an impulsive expansion consequent to the rapid 
gas mass loss. The former includes the slow stellar mass loss associated with the death 
of old stars, i.e., AGB winds. In addition, the slow expulsion of gas during the quiescent 
phases of AGN can also adiabatically expand the central region. The latter, rapid mass 
loss involves SN driven winds as well as AGN driven winds \citep{2008ApJ...689L.101F}, 
with a shorter ejection timescale then the dynamical timescale. 

If we define the fractional 
changes in mass and in radius of a self gravitating system, as 
$\delta_m \equiv (m_1 - m_0) / m_0$, and $\delta_r \equiv (r_1 - r_0) / r_0$, where 
$m_0$ and $m_1$ are the initial and final masses after mass loss and
$r_0$ and $r_1$ are the initial and final radii, we have adiabatic expansion with 
a rate of $\delta_r = - \delta_m / (\delta_m + 1)$ after mass loss on a timescale longer
than a dynamical timescale. But when the gas mass is ejected with a timescale shorter
than the dynamical timescale of the system, the expansion proceeds at
a higher rate as $\delta_r = - \delta_m / (2\delta_m + 1)$. For example, for the 25 percent
of radius increase, $\delta_r = 0.25$, which leads to $\sim 50$ percent decrease in
density, we require a mass loss of $\delta_m = 1/3$ for the adiabatic case, but only
$\delta_m = 1/6$ for the impulsive mass loss case. Outflowing gas in our galaxies simulated 
with AGN driven wind feedback show much higher characteristic velocity 
($500-1000 \kms$) \citep{2018ApJ...860...14B}.

Finally, core scouring by binary black hole systems can heat and expel collisionless 
matter from the central region the gravitational slingshot effect 
\citep{2001ApJ...563...34M}. The black hole scouring effect is able to produce the observed 
light profiles of galaxy cores \citep{1997AJ....114.1771F}, and its impact is found to be
imprinted in recently observed massive elliptical galaxies \citep{2016Natur.532..340T},
which show a tight correlation between the sizes of the core in the observed light
profile and the radii of the black hole's sphere of influence.  Our simulation
incorporates black hole mergers and naturally includes the heating effect of the
corresponding black hole binary orbital decays. However, the resolution in our
simulation is not sufficient to capture this effect of black hole scouring in detail, 
as the black hole binary scouring is considered to be mainly important on length 
scales of $\le 100$ pc \citep[see][]{2017ApJ...840...53R,2018arXiv180510295R}.

In order to check the contribution of the ``puffing-up'' process to the stellar core decrease,
we show the histogram of stellar radial migration distance from $z=2$ to $z=0$ for 
all star particles which were within the central region $r <1$ kpc at $z=2$ in 30 WithAGN galaxies
in Figure~\ref{fig:puff}. 
Although there are a few star particles that travelled inward
(radial travel distance $< 0$), most of stars migrate radially outwards since $z=2$.
Over 70 percent of star particles migrate radially outward more than 1 kpc since 
$z=2$, that is, the majority of star particles which constituted the core at $z=2$ are 
no longer within the central region. This is primarily due to the expansion after gas ejection as discussed
above. \citet{2017MNRAS.465..722F} also recently found such a migration
of stars in their compact galaxies selected from the EAGLE simulation. They showed that
this star migration from the central region of galaxies can result in the size growth 
of compact galaxies, but its contribution is modest, when renewed star formation 
and mergers dominate the size growth.

\subsection{Evolution of galaxies in the sSFR-$\Sigma_{\rm eff} $ plane}


In order to study the structural evolution of galaxies and its connection
to quenching of star formation, in Figure~\ref{fig:sSFR_sigma} we show 
the evolution of simulated central galaxies in the plane of specific star 
formation rate (sSFR) versus the effective stellar density 
$\Sigma_{\rm eff} (= M/\pi \reff^{2})$, following
\citet{2013ApJ...765..104B}. In the left panels of 
Figure~\ref{fig:sSFR_sigma}, we show sSFR and $\Sigma_{\rm eff}$ of 
the 30 central galaxies simulated with and without AGN feedback at 
$z=0.5$, and their most massive progenitors at $z=1,2$ and $3$. In the 
right panel of Figure~\ref{fig:sSFR_sigma}, we show the average tracks 
of simulated galaxies in sSFR-$\Sigma_{\rm eff}$ plane for the 
WithAGN and NoAGN models, showing the schematic evolutionary 
tracks with the arrows. At $z=3$, all progenitor galaxies simulated with 
and without AGN feedback are distributed horizontally in the 
sSFR-$\Sigma_{\rm eff}$ plane, with almost constant sSFR over a 
broad range in $\Sigma_{\rm eff}$. By $z=2$ galaxies in both models 
move toward the right, becoming more compact. Note that we see this 
``compaction'' trend even before galaxies start to quench, presumably 
due to dissipative processes such as gas rich mergers and disk 
instabilities \citep{2015MNRAS.450.2327Z,2016MNRAS.458..242T}. 
By $z=0.5$, the galaxies simulated with AGN feedback show a
rapid decline in sSFR, moving downwards in the diagram rapidly, as 
the quenching time scale is very short 
\citep[see][]{2017MNRAS.472.2054P}.  NoAGN galaxies also show a 
decline in their sSFR, but the changes are much smaller and slower, 
and they do not decline below $\rm log sSFR=-1.5$. After star 
formation is quenched, galaxies simulated with AGN decrease their 
stellar densities, moving toward the left on the sSFR-$\Sigma_{\rm eff}$ 
plane, while galaxies simulated without AGN feedback show a relatively 
small decrease in their stellar density.

We note that galaxies simulated with AGN feedback evolve along a strikingly similar evolutionary 
track to the one suggested by \citet[][see their Figure 6]{2013ApJ...765..104B}. They characterized the structural 
evolution of galaxies in the $\Sig15 (= M/\reff^{1.5})$ plane, and showed that
galaxies follow three sequences, (1) star-forming and diffuse galaxies transformed
into compact star forming galaxies via  multiple dissipative processes, (2) compact
star forming galaxies turned into compact quiescent galaxies by rapid quenching 
of star formation, and finally, (3) in compact quiescent galaxies the galaxy sizes are extended via 
the gradual growth and expansion due to multiple dry mergers.
The galaxies simulated with AGN feedback show a similar `clockwise' evolutionary 
track and similar transitions from extended SFGs, to compact SFGs, to compact QGs, and
finally to extended QGs. In contrast, the galaxies in the NoAGN simulations show much less dramatic evolution.

Figure~\ref{fig:track1} shows the evolution of two galaxies: we show the
evolution tracks in sSFR versus $\Sigma_{\rm eff}$ in the center panel
and their projected stellar density snapshots at redshift $z=3,2,1$
and $z=0$. Again, the galaxies follow similar tracks to those shown in \citet{2013ApJ...765..104B}.
The galaxy sizes become smaller at a roughly constant specific 
star formation rate, and the onset of quenching again occurs at a nearly
constant $\Sigma_{\rm eff}$.

\section{Summary and Discussion}\label{discussion}
We have explored the role of AGN feedback on the evolution of galaxy 
sizes, compactness and core densities using numerical cosmological hydrodynamical 
simulations with an observationally based novel sub-grid treatment of 
winds driven by radiatively efficient accretion onto super massive black hole. The results
can be summarized as follows.

\begin{itemize}
\item We show that galaxies
simulated with AGN feedback follow much steeper evolution with 
$\Delta r\propto\Delta M^{2}$ in the size-mass plane compared to
galaxies without AGN feedback (Figure~\ref{fig:r_vs_mstar}),
and show rapid size evolution with redshift, $\reff \propto (1+z)^{-1.49}$ (Figure~\ref{fig:r_vs_z}).

\item AGN feedback enhances the ``inside out'' evolution of 
massive galaxies via efficient star formation quenching mechanisms 
\citep[e.g.][]{2010MNRAS.401.1099H,2013ApJ...771...85V}, by effectively
quenching in-situ star formation therefore increasing the fraction of 
accreted stars in a galaxy (Figure~\ref{fig:r_vs_fraction}). 

\item We also show that galaxies simulated with AGN feedback 
decrease their core densities after the quenching of star formation, 
while galaxies without AGN feedback continue to increase their core 
densities as well as stellar mass (Figure~\ref{fig:sigma1_vs_mstar}). 
The decrease in core density is caused by stellar mass loss
as well as ``puffing-up'' process in response to gas mass loss (Figure~\ref{fig:puff}).
\end{itemize}

We also study the evolution of galaxies in the star formation versus 
compactness (size) plane (Figure~\ref{fig:sSFR_sigma}).
We show the role of AGN feedback
on the formation of compact quiescent galaxies at $z\sim2$ and present-day
extended quiescent galaxies through the evolutionary 
scenario suggested by \cite{2013ApJ...765..104B}, which can be summarized as follows.

\begin{itemize}
\item[1.] The formation of compact quiescent
galaxies first follows an evolutionary sequence from
extended star forming galaxies (upper-left) to compact star forming galaxies (upper-right)
due to a dissipative process, such as highly dissipative mergers 
between gas-rich progenitors, which are more common at 
high redshift \citep[e.g.][]{2011MNRAS.415.3135C}. 

\item[2.] Then the AGN feedback quenches the star formation of compact star forming galaxies 
 and lowers their sSFR at roughly constant $\Sigma_{\rm eff}$, populating
the compact quiescent galaxy region (lower-right) rapidly.

\item[3.] At later times, after in-situ star formation has been quenched, 
the simulated galaxies increase their stellar mass primarily 
by the accretion of smaller stellar systems, which 
leads to a strong size growth 
\citep[][but see \citealp{2012ApJ...746..162N,
2012MNRAS.422.1714N}]{2006ApJ...636L..81N,2009ApJ...699L.178N,
2012ApJ...744...63O,2012ApJ...754..115J,
2012MNRAS.425.3119H,2013MNRAS.428..641O,
2013MNRAS.429.2924H,2013MNRAS.431..767B}.
Minor mergers predominantly deposit materials in the 
outskirts of the galaxies and finally form extended 
quiescent galaxies populating lower-left region of 
sSFR-$\Sigma_{\rm eff}$ plane.
\end{itemize}

In summary, mechanical AGN feedback plays an important role in making massive galaxies red and 
dead, and it also plays an important role in making them extended. Including mechanical 
AGN feedback enhances the size growth in different ways. First of all, it has indirect, 
supporting but major impact, as with effective in-situ star formation quenching, it increases 
the fraction of accreted stars \citep{2012MNRAS.419.3200H,2013MNRAS.433.3297D}, 
so the impact of dry mergers is much enhanced. Moreover, since 
completely dry mergers are less frequent due to the high gas fractions in satellite galaxies without AGN feedback, 
including AGN feedback can increase the frequency of dry mergers. Secondly, it 
has direct but probably minor impact, as AGN-driven gas outflows can produce 
fluctuations of the gravitational potential  and puff-up the central region. It will also 
enhance the adiabatic expansion in response to the slow expulsion of gas from the 
central region during the quiescent mode of AGN activity.

\begin{acknowledgements}
The authors would like to  thank the anonymous referee for helpful comments.
We also thank Guillermo Barro, Renyue Cen, Sandra Faber, Shy Genel, and 
Sarah Wellons for helpful conversations and suggestions. 
Numerical simulations were run on the computer clusters of the Princeton 
Institute of Computational Science and engineering. This research was 
supported by NASA through grant number HST Cycle 23 AR-14287 from 
the Space Telescope Science Institute.
 
\end{acknowledgements}

\bibliography{references}


\end{document}